\documentclass[conference]{IEEEtran}
\usepackage{cite}
\usepackage{amsmath,amssymb,amsfonts}
\usepackage{algpseudocode}
\usepackage{graphicx}
\usepackage{textcomp}
\usepackage{xcolor}
\usepackage{hyperref}
\usepackage{tikz}
\usepackage{color,soul}
\usepackage{tabularx}
\usepackage{booktabs}
\usepackage{multirow}
\usepackage{subcaption}
\usepackage{makecell}
\usepackage{bbm}
\usepackage{algorithm}

\usetikzlibrary{quantikz2}
\usetikzlibrary{positioning, shapes.geometric, arrows.meta, calc}
\def\BibTeX{{\rm B\kern-.05em{\sc i\kern-.025em b}\kern-.08em
    T\kern-.1667em\lower.7ex\hbox{E}\kern-.125emX}}

\newcolumntype{?}{!{\vrule width 1.5pt}}

\hypersetup{
    colorlinks=true,
    linkcolor=blue,
    filecolor=magenta,      
    urlcolor=cyan,
    citecolor=blue,
}

\usepackage{changes}
\definechangesauthor[name={kien}, color=blue]{kn}

\begin{document}

\title{Q3SAT-GPT: A Generative Model for Discovering Quantum Circuits for the 3-SAT Problem
}

    


\author{
\IEEEauthorblockN{
Pratim Ugale\IEEEauthorrefmark{1},
Ilya Tyagin\IEEEauthorrefmark{1},
Karunya Shirali\IEEEauthorrefmark{2}\IEEEauthorrefmark{3},
Kien X. Nguyen\IEEEauthorrefmark{1},
Ilya Safro\IEEEauthorrefmark{1}\IEEEauthorrefmark{4}
}
\IEEEauthorblockA{\IEEEauthorrefmark{1}\textit{Department of Computer and Information Sciences, University of Delaware}, Newark, DE, USA}
\IEEEauthorblockA{\IEEEauthorrefmark{2}\textit{Department of Physics, Virginia Tech}, Blacksburg, VA, USA}
\IEEEauthorblockA{\IEEEauthorrefmark{3}\textit{Virginia Tech Center for Quantum Information Science and Engineering}, Blacksburg, VA, USA}
\IEEEauthorblockA{\IEEEauthorrefmark{4}\textit{Department of Physics and Astronomy, University of Delaware}, Newark, DE, USA}
\IEEEauthorblockA{\{pratim, tyagin, kxnguyen, isafro\}@udel.edu, karunyashirali@vt.edu}
}

\maketitle

\begin{abstract}
This work introduces Q3SAT-GPT, a generative model for discovering quantum circuits for the Max-E3-SAT problem. Our method learns from high-performing QAOA-style ansätze to directly generate candidate circuits. To create high-quality supervision, we also introduce Mosaic Adaptive QAOA (MosaicADAPT-QAOA), an adaptive strategy for constructing low-depth QAOA circuits by selecting subsets of mixer operators in each step, rather than inserting operators sequentially. The resulting circuits serve as training data for the generative model, allowing it to learn effective circuit design patterns while eliminating the need for costly variational optimization at inference time. Experiments show that our framework attains strong solution quality with shallow circuits and scales significantly better than both our adaptive construction procedure and conventional variational baselines. Our results establish generative modeling as a high-performance route toward the scalable discovery of quantum optimization circuits, demonstrating that these models can effectively internalize circuit logic while providing a foundation for future, instance-aware inductive biases. Reproducibility: The source code is available at \url{https://github.com/pratimugale/Q3SAT-GPT}.
\end{abstract}

\begin{IEEEkeywords}
Q3SAT-GPT, MosaicADAPT-QAOA, QAOA, Max3SAT, QAOA-GPT, Quantum Optimization, Generative AI
\end{IEEEkeywords}

\section{Introduction}


Quantum computing faces a central scientific bottleneck: useful fault-tolerant machines will not deliver impact through hardware alone, but through the discovery of new algorithms that exploit problem structure (without which possible  complexity improvements are severely limited \cite{bennett1997strengths,fortnow1999complexity,aaronson2022much}) while respecting such resource constraints as logical qubits, circuit depth, and error-corrected compilation \cite{lougovski2023report,beverland2022assessing,katabarwa2024early}. Quantum algorithms for combinatorial optimization are not an exception. While application domains for it are broad spanning finance \cite{Herman_2023}, network science \cite{shaydulin2019network}, and computational biology \cite{fedorov2021towards} to mention just a few, at present, most quantum algorithms for optimization are hand-designed, or adapted from generic templates. This is especially limiting for combinatorial problems, where naive encodings often destroy structure, inflate resource counts, and erase any plausible advantage. The critical challenge, therefore, is not merely to run known quantum routines on better hardware, but to create AI systems that can discover reduced-complexity, fault-tolerant quantum algorithms tailored to the problems of interest, together with compilation and verification tools that make those algorithms executable and trustworthy \cite{alexeev2025artificial,merilehto2026generative}. 

The goal of this work is to design a generative AI framework for one of the most broadly applicable NP-hard optimization problems: Max-E3-SAT \cite{haastad2001some} (the problem of maximizing the number of satisfied clauses in CNF-SAT instances with exactly three literals in each clause). Max-E3-SAT (also referred to as 3-SAT) is tightly connected to constraint satisfaction and has motivated decades of research in both exact and approximate classical optimization, resulting in a large ecosystem \cite{SATCompetitions,jarvisalo2012international} of highly engineered solvers and benchmarks. This makes it an especially meaningful testbed for generative quantum algorithm design: any proposed quantum-circuit synthesis method must be evaluated against a problem class for which classical baselines are  strong, and extensively studied. Another aspect of interest is that, similarly to MaxCut (a traditional target for quantum combinatorial optimization due to its direct connection to the Ising model), Max-E3-SAT guarantees simple but strong approximation ratios which are by themselves not bad, e.g., 
H{\aa}stad showed that, unless $\mathrm{P}=\mathrm{NP}$, no polynomial algorithm can guarantee an approximation ratio better than $7/8+\varepsilon$ for any fixed $\varepsilon>0$ \cite{haastad2001some}. 
Thus, Max-E3-SAT provides an excellent setting for studying whether generative AI can discover problem-aware quantum circuit structures that go beyond templates and well known approximations.

The Quantum Approximate Optimization Algorithm (QAOA)~\cite{farhi2014quantum} has emerged as a leading variational framework for combinatorial optimization, with the potential to provide advantages for certain problem classes or instance families. Its near-term relevance stems from its hybrid quantum-classical structure: a parameterized quantum circuit prepares candidate solution states, while a classical optimizer updates the variational parameters. This architecture makes QAOA particularly attractive for both noisy intermediate-scale quantum and fault-tolerant regimes. However, original QAOA and most of its subsequent versions \cite{abbas2024challenges} have a rigid circuit structure, which directly impacts expressibility and  inability to cope with barren plateaus and can substantially degrade solution quality  before circuits reach depths sufficient to exploit problem structure.

Several variants of QAOA have been proposed to improve different aspects of the original algorithm. Parameter transferability and initialization strategies aim to reduce the classical optimization overhead by exploiting concentration and transferability of good angles across related instances~\cite{falla2024graph,brandao2018fixed, galda2021transferability}. Warm-start QAOA incorporates solutions of classical relaxations into the initial quantum state~\cite{egger2021warm}. Multi-angle QAOA increases expressivity by assigning independent parameters to different terms in the cost and mixer Hamiltonians~\cite{herrman2022multi}. Sparsification of the phase operator reduces the number of gates in QAOA \cite{liu2022quantum}. Recursive QAOA combines variational optimization with iterative variable elimination~\cite{bravyi2022hybrid}. In this sense, QAOA is not merely a quantum algorithm, but a hybrid optimization paradigm whose success depends on carefully balancing quantum expressivity with classical control.

The most relevant to this work are adaptive approaches that change the circuit structure \cite{adapt-qaoa,liu2022layer,magann2022feedback}. In particular, ADAPT-QAOA constructs the ansatz problem-dependently by selecting operators from a pool according to gradient information. More recent extension, TETRIS-ADAPT further seeks to reduce circuit depth and optimization cost by selecting multiple  operators per layer~\cite{anastasiou2024tetris}. 

In this work, we develop a GPT-based generative approach for discovering quantum algorithms for Max-E3-SAT. Rather than treating quantum circuit design as a purely hand-crafted or repeatedly optimized variational task, our framework learns circuit generation patterns from high-quality adaptive circuits and then produces candidate circuits directly by autoregressive generation. To construct the training corpus, we introduce MosaicADAPT-QAOA, a modification of the TETRIS-ADAPT strategy that selects compatible sets of mixer operators at each layer \emph{by optimizing their collective contribution}, rather than selecting the set of mixer operators to add at each layer by greedily inserting operators one at a time as in TETRIS-ADAPT. This produces shallow, problem-adapted circuits that serve as supervision for the generative model. The resulting Q3SAT-GPT framework is therefore designed to amortize the cost of slow adaptive circuit construction and variational parameter optimization: the expensive MosaicADAPT-QAOA procedure is used offline to generate training data, while inference requires only a single forward generative pass to propose a quantum circuit for a new Max-E3-SAT instance. 


Another challenge in QAOA is the optimization of the parameters in the ansatz. The optimization involves measuring the expectation energy of the circuit, and updating the parameters to optimize the expectation. While adaptive methods lead to shallower circuits by constructing problem-tailored ans\"{a}tze, repeated energy gradient measurements are a significant overhead to the computational cost of the algorithm. Even after selecting the best operator, the parameters of the quantum circuit must again be optimized, along with the parameters associated with the newly added operator. Similar to our previous work QAOA-GPT \cite{tyagin2025qaoa}, in Q3SAT-GPT we also generate fully optimized quantum circuits end-to-end. 


In summary, our contributions are:

\begin{enumerate}
    \item \textbf{MosaicADAPT-QAOA:} The dense tiling method that extends ADAPT and TETRIS methods. The tiling is designed to  optimize the sum of gradients of disjoint operators. For example, we observe that MosaicADAPT-QAOA requires a median of 4 fewer layers to reach 99.9\% approximation ratio on 10 variable problems.
\item \textbf{Q3SAT-GPT:} We introduce a GPT-based generative framework for synthesizing quantum circuits for Max-E3-SAT instances. Using high-quality circuits produced by MosaicADAPT-QAOA as supervision, the model learns structural and parametric patterns that characterize effective adaptive ans\"{a}tze for this problem class. Once trained, Q3SAT-GPT generates candidate circuits in a single autoregressive inference pass. This positions Q3SAT-GPT as a scalable AI generative approach for quantum circuit discovery, providing strong generated circuits for Max-E3-SAT while opening a path toward instance-aware generative design of quantum optimization algorithms.    
\end{enumerate}

\section{Background}

\subsection{The Max-E3-SAT Problem}
The Max-E3-SAT problem \cite{JOHNSON1974256} is an NP-hard combinatorial optimization problem \cite{PAPADIMITRIOU1991425}. The goal is to maximise the number of satisfied clauses in a formula in its Conjunctive Normal Form (CNF), where each clause comprises of a disjunction (logical OR) of exactly three literals. 

Let $\mathcal{F} = C_1 \land \dots \land C_m$ be a CNF formula containing $m$ clauses over $n$ Boolean variables $x_i \in \{0, 1\}$. A literal $l$ is defined as either a variable $x_i$ or its negation $\neg x_i$. A clause $C_r$ is represented as:
$$C_r = (l_{r,1} \lor l_{r,2} \lor l_{r,3}),$$
and is satisfied if at least one of its constituent literals evaluates to $1$ (True). The goal of a Max-E3-SAT solver is to find a variable assignment that maximizes:$$\sum_{r=1}^{m}\mathbbm{1}[C_r \text{ is satisfied}],$$ where $\mathbbm{1}[\cdot]$ is the indicator function.
A simple randomized algorithm that independently assigns truth values to variables with a probability of $1/2$ satisfies an expected $87.5\%$ of the clauses, as each clause in a 3-CNF formula has a $7/8$ probability of containing at least one true literal. This random baseline is useful when interpreting approximation ratios, especially for satisfiable instances where the optimum equals the total number of clauses.

\noindent {\bf Uniform Random Formulas}.
The Uniform Random 3-SAT model defines a specific distribution of CNF formulas \cite{hoos2000satlib}. In this approach, a formula is constructed by independently generating $m$ clauses over $2n$ literals, while discarding clauses with redundant literals or contradictory variable pairs.

\noindent {\bf Balanced Formulas. }
The generation protocol for balanced Max-E3-SAT instances strictly enforces variable regularity, which means that the total frequency of each variable across the formula varies by no more than a single occurrence ($\pm 1$) \cite{spence2017balanced}. This rule also applies to literal polarity, providing an almost perfectly equal split between positive variables and their negations, while minimizing structural overlap between clauses.

\noindent {\bf Critical Phase Transition Ratio. }
The difficulty and satisfiability of Max-E3-SAT instances is governed by its phase transition ratio, defined as the ratio of number of clauses in formula to the number of distinct variables $\alpha = m/n$. The probability of an instance being satisfiable undergoes a sharp decline as the clause count $m$ for a given number of variables $n$ increases toward the critical phase transition ratio \cite{mitchell1992hard, hoos2000satlib}. For Uniform Random 3-SAT instances, the value $\alpha \approx 4.26$ marks the critical phase transition between satisfiable and unsatisfiable regimes \cite{hoos2000satlib}, while for Balanced instances, this phase transition ratio $\alpha$ is found to be at $3.6$ \cite{spence2017balanced}.

\subsection{The Quantum Approximate Optimization Algorithm}
The Quantum Approximate Optimization Algorithm (QAOA) \cite{qaoa} aims to find a bitstring solution  $x \in \{0,1\}^n$ that optimizes a classical function. 
The classical objective function is first encoded into a cost Hamiltonian $H_C$, typically formulated as an Ising model consisting of Pauli-$Z$ operators. The algorithm initializes n qubits in a uniform superposition $|+\rangle^{\otimes n}$.
The parameterized QAOA ansatz is constructed by applying $p$ alternating layers of the cost Hamiltonian $H_C$ and a mixing Hamiltonian $H_M$, parameterized by $2p$ angles $\boldsymbol{\gamma} = \{\gamma_i\}^p_{i=1}$ and $\boldsymbol{\beta} = \{\beta_i\}^p_{i=1}$:
$$|\psi(\boldsymbol{\gamma}, \boldsymbol{\beta})\rangle = \prod_{k=1}^p \left( e^{-i\beta_k H_M} e^{-i\gamma_k H_C} \right) |s\rangle$$

To optimize the state, a classical optimizer iteratively updates $\boldsymbol{\gamma}$ and $\boldsymbol{\beta}$ to minimize the expectation value of the cost Hamiltonian:$$E(\boldsymbol{\gamma}, \boldsymbol{\beta}) = \langle \psi(\boldsymbol{\gamma}, \boldsymbol{\beta}) | H_C | \psi(\boldsymbol{\gamma}, \boldsymbol{\beta}) \rangle$$

Achieving high quality solutions with QAOA typically require a high number of layers, which is challenging to execute on current quantum hardware. Previous studies have proposed parameter-transfer and initialization strategies to reduce the optimization time by exploiting concentration and transferability of good angles across related instances~\cite{brandao2018fixedcontrolparametersquantum,galda2021transferability,galdasimilarity}.

\subsection{Adaptive Methods for Ansätze Creation}
Methods such as ADAPT-VQE \cite{adaptvqe} and ADAPT-QAOA \cite{adapt-qaoa} address the limitations of current devices and the locality constraints of standard QAOA \cite{bravyi2020obstacles} by adaptively and iteratively building ansätze tailored to the specific problem. By selecting only the most impactful operators, these algorithms achieve accelerated convergence. ADAPT-QAOA replaces the standard fixed mixer in QAOA with a pool of operators $\mathcal{P}$. At each layer of QAOA, the algorithm selects one operator $\hat{A}_j$ from this pool based on an adaptive selection process. The selected operator is the one whose addition to the circuit is estimated to produce the largest immediate improvement in the QAOA objective. 

In contrast to the standard QAOA mixer which acts on all qubits, the operator pool in ADAPT-QAOA consists of single-qubit and two-qubit Pauli operators. The operator selection process for ADAPT-QAOA~\cite{adapt-qaoa} was originally inspired by ADAPT-VQE \cite{adaptvqe}, which selects a single operator per layer. TETRIS-ADAPT-VQE \cite{anastasiou2024tetris} is a recent development that allows the selection of multiple disjoint operators in a single step and has been shown to result in significantly shallower circuits compared to ADAPT-VQE. This is because multiple operators can be added to the ansatz at the same step, allowing the expressivity to grow while preventing any qubit from sitting idle at a particular layer.

\subsection{QAOA-GPT}

While algorithmic variations like ADAPT-QAOA dynamically tailor the ansatz to the problem instance, the iterative evaluation of gradients at each layer, along with the classical optimization of the variational parameters, incurs a significant classical computational overhead. To address this bottleneck, recent literature has explored data-driven, machine learning frameworks. QAOA-GPT \cite{tyagin2025qaoa} is a generative pre-trained transformer model designed to synthesize quantum circuits in a single inference step. 

QAOA-GPT was trained on a large dataset of graph instances paired with their corresponding optimized circuits generated via standard ADAPT-QAOA. The model bypasses the iterative gradient calculations and parameter optimization entirely. While such generative approaches offer massive speedups at inference time, their performance is inherently bounded by the quality of the training data. Therefore, advancing the underlying adaptive algorithms that generate this training data, remains a critical necessity for the scalable success of AI-driven quantum circuit generation.

\section{Methodology}

\subsection{Overview of the Pipeline}

The proposed methodology has two stages.
First, MosaicADAPT-QAOA constructs high-quality adaptive QAOA circuits for Max-E3-SAT by selecting compatible sets of mixer operators at each layer.
These circuits are then used as supervision for Q3SAT-GPT, which conditions on an input 3-CNF formula and its Literal-Clause Graph (LCG) embeddings to generate a tokenized MosaicADAPT-QAOA-style circuit in a single autoregressive pass.

\subsection{Problem Formulation and Hamiltonian Encoding}
To evaluate the algorithm's performance on unapproximated problem dynamics, we construct an exact cost Hamiltonian for Max-E3-SAT using a Higher-Order Unconstrained Binary Optimization (HUBO) formulation similar to \cite{exact_hamiltonian_for_max3sat}.
We map the classical binary variables $x_i \in \{0, 1\}$ to Pauli-$Z$ operators via the transformation $x_i = \frac{I - Z_i}{2}$, where the computational basis state $|0\rangle$ corresponds to a logical $0$. For each clause $C_r$, we construct a penalty term $H_{C_r}$ by taking the product of the literal penalties. If the $j$-th literal in the clause is a positive variable $x_i$, its penalty is $(1 - x_i)$; if it is a negative literal $\neg x_i$, its penalty is $x_i$. Mapping these to Pauli-$Z$ operators, the clause Hamiltonian becomes:
\begin{equation}
H_{C_r} = \prod_{j=1}^{3} P_{r,j}, \quad \text{where } P_{r,j} =
\begin{cases}
\frac{I + Z_{i}}{2} & \text{if } l_{r,j} = x_i \\
\frac{I - Z_{i}}{2} & \text{if } l_{r,j} = \neg x_i.
\end{cases}
\end{equation}
The total cost Hamiltonian is constructed by summing these terms over all $m$ clauses: $H_C = \sum_{r=1}^m H_{C_r}$. Consequently, the expectation value $\langle H_C \rangle$ corresponds directly to the expected number of violated clauses, and the ground state energy $E_0$ represents the global minimum of unsatisfied clauses.

\subsection{Operator Pool for Max-E3-SAT}\label{sec:operator_pool}



The ADAPT-QAOA framework requires an operator pool $\mathcal{P}$ from which to choose operators to form each mixer layer of the ansatz. 
The original ADAPT-QAOA pool included only operators that contained an even number of $Y$ or $Z$ Paulis as the problem exhibits bit-flip symmetry.  The native formulation of the Max-E3-SAT problem does not exhibit this bit-flip symmetry. 
We construct an expanded pool containing all combinations of the standard QAOA mixer, and single and two-qubit Pauli operators. 
Single qubit $Z$ and two-qubit $ZZ$ operators are excluded from the pool, as they commute with the Cost Hamiltonian $H_C$ comprising of only $Z, ZZ$ and $ZZZ$ type operators. 
Our complete operator pool $\mathcal{P}$ is therefore defined as:
\begin{equation}
\begin{split}
    \mathcal{P} = {} & \left\{ \sum_{k=1}^n X_k \right\} \cup \{ X_i, Y_i \}_{\forall i} \\
    & \cup \{ X_i X_j, Y_i Y_j, X_i Y_j, X_i Z_j, Y_i Z_j \}_{\forall i, j, \, i \neq j}.
\end{split}
\end{equation}

\subsection{MosaicADAPT-QAOA for High Quality Supervision}
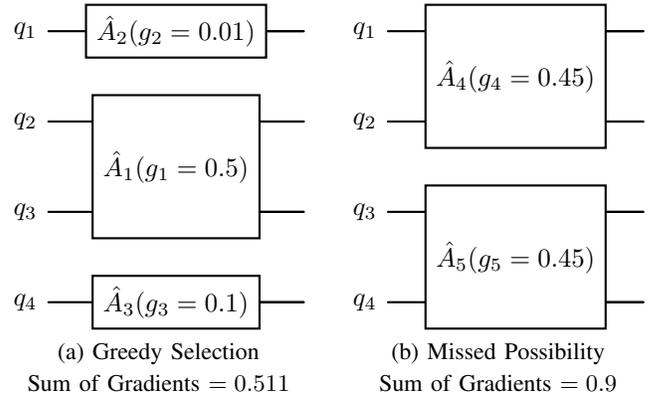
\begin{figure}[h]
    \centering
    \begin{tabular}{cc}
        \begin{quantikz}
            \lstick{$q_1$} & \gate{\hat{A}_2 (g_2=0.01)} & \qw \\
            \lstick{$q_2$} & \gate[2]{\hat{A}_1 (g_1=0.5)} & \qw \\
            \lstick{$q_3$} &                     & \qw \\
            \lstick{$q_4$} & \gate{\hat{A}_3 (g_3=0.1)} & \qw 
        \end{quantikz} & 
        \begin{quantikz}
            \lstick{$q_1$} & \gate[2]{\hat{A}_4 (g_4=0.45)} & \qw \\
            \lstick{$q_2$} &                      & \qw \\
            \lstick{$q_3$} & \gate[2]{\hat{A}_5 (g_5=0.45)} & \qw \\
            \lstick{$q_4$} &                      & \qw 
        \end{quantikz} \\
        \small (a) Greedy Selection & \small (b) Missed Possibility \\
        \small Sum of Gradients $= 0.511$ & \small Sum of Gradients $= 0.9$
    \end{tabular}
    \caption[A Representative Conflict Illustration.]{A Representative Conflict Illustration. Operator generators $\hat{A}_i$ are labeled with their corresponding gradient magnitudes in parentheses. In (a), selecting the highest individual gradient ($\hat{A}_1$) blocks the selection possibility of $\hat{A}_4$ and $\hat{A}_5$, leading to a lower total sum of gradients as compared to (b).}
    \label{fig:tiling_conflict}
\end{figure}

We observe that the operator selection method in TETRIS-ADAPT-VQE is based on a greedy strategy, as it chooses the operator with the current best gradient magnitude. It is susceptible to cases where selecting the current best operator prevents the inclusion of multiple near-best operators that in combination could have provided a greater total energy reduction. See Figure \ref{fig:tiling_conflict} for a representative example.

\begin{figure}[h]
    \centering
    \includegraphics[width=0.9\columnwidth]{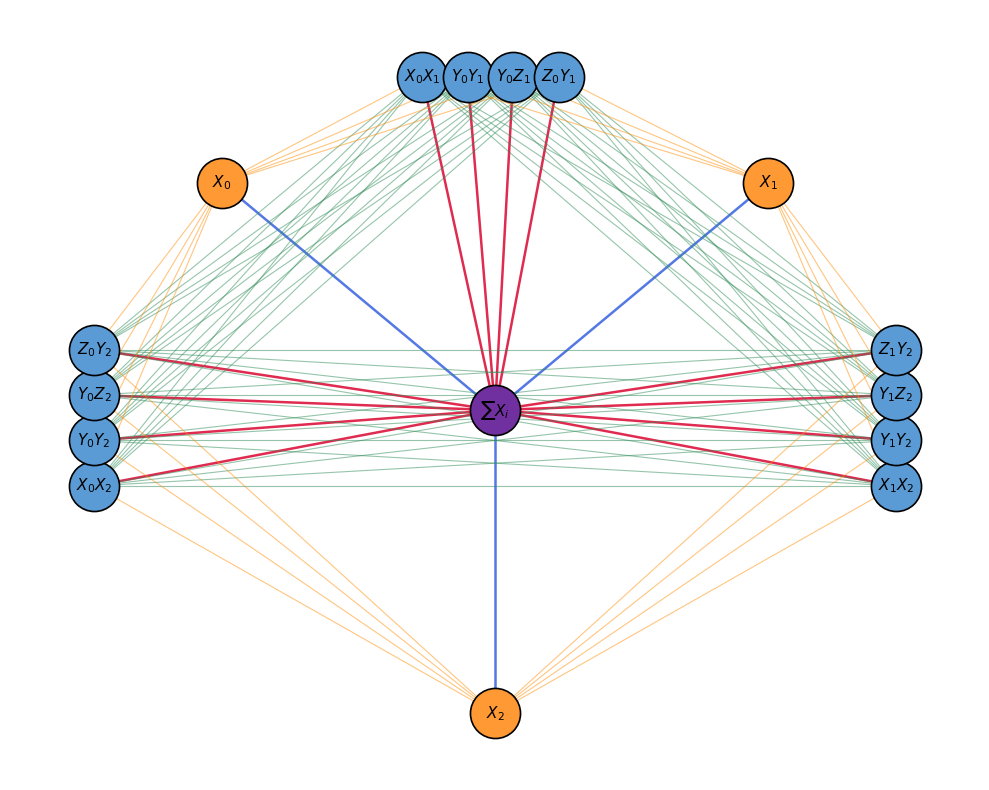}
    \caption[Incompatibility graph representation for a 3-qubit problem instance.]{Incompatibility graph representation for a 3-qubit problem instance, used for operator selection. Each node in the graph represents a candidate unitary operator from the pool $\mathcal{P}$, An edge connects two nodes if and only if their corresponding operators have overlapping qubit support. Sets of overlapping two-qubit operators form fully connected cliques and are omitted to maintain legibility.}
    \label{fig:incompatibilitygraph}
\end{figure}

We propose a selection criterion that aims to maximize the global sum of gradient magnitudes for all selected operators at each QAOA layer. Formally, we model disjoint operator selection at a particular layer as a Maximum Weight Independent Set (MWIS) problem on an incompatibility graph. In such a graph, the nodes represent operators, the node weights are kept proportional to their respective gradient magnitudes, and the edges denote overlapping qubit supports (Figure \ref{fig:incompatibilitygraph}). Solving MWIS on this graph will yield a disjoint set of operators whose sum of gradients is maximized.
Since MWIS is NP-hard, we employ an approximate classical approach KaMIS  \cite{DBLP:journals/heuristics/LammSSSW17}. Specifically, we utilize the Memetic MWIS solver \cite{dblp:conf/gecco/grossmannl0s23}, along with its internal scalable kernelization routines \cite{DBLP:journals/jea/Hespe0S19}, to maximize the total gradient sum per layer.

Given an operator pool $\mathcal{P}$, cost Hamiltonian $\hat{H}_C$, initial cost-evolution angle $\gamma_0$, gradient threshold $\epsilon_g$, and maximum number of adaptive layers $L_{\max}$, MosaicADAPT-QAOA initializes the state in the uniform superposition and iteratively adds one mixer layer at a time.
At each iteration $k$, the previous state $|\psi^{(k-1)}\rangle$ is first evolved by $e^{-i\gamma_0\hat{H}_C}$ for gradient evaluation.
For every candidate operator $\hat{A}_j \in \mathcal{P}$, we compute the gradient score $g_j = -i \langle \psi^{(k-1)} | e^{-i\gamma_0\hat{H}_C}[\hat{H}_C, \hat{A}_j] e^{-i\gamma_0\hat{H}_C} | \psi^{(k-1)} \rangle$.
The candidate operators are then represented as an incompatibility graph whose edges connect operators with overlapping qubit support.
Solving the resulting MWIS instance gives the disjoint operator set $\mathcal{A}_{\mathrm{layer}}$ added at the next mixer layer.
After each addition, all variational parameters $(\boldsymbol{\gamma},\boldsymbol{\beta})$ are reoptimized with BFGS~\cite{nocedal.wright:99}.
The procedure stops when all gradient scores fall below $\epsilon_g$ or when $L_{\max}$ layers have been added.
The overall MosaicADAPT-QAOA procedure is described at a high level in Algorithm \ref{alg:MosaicADAPT-QAOA}.
\begin{algorithm}[htbp]
\small
\caption{MosaicADAPT-QAOA}
\label{alg:MosaicADAPT-QAOA}
\begin{algorithmic}[1]
\Require $\mathcal{P}$, $\hat{H}_C$, $\gamma_0$, $\epsilon_g$, $L_{\max}$
\Ensure Optimized parameters $(\boldsymbol{\gamma}, \boldsymbol{\beta})$ and final state $|\psi\rangle$
\State $k\gets0$
\State $|\psi^{(0)}\rangle \gets |+\rangle^{\otimes n}$; $\boldsymbol{\gamma},\boldsymbol{\beta}\gets\emptyset$
\While{$k < L_{\max}$}
    \For{$\hat{A}_j \in \mathcal{P}$}
        \State $g_j = -i \langle \psi^{(k-1)} | e^{-i\gamma_0\hat{H}_C}[\hat{H}_C, \hat{A}_j] e^{-i\gamma_0\hat{H}_C} | \psi^{(k-1)} \rangle$
    \EndFor
    \If{$\max_j \lVert g_j \rVert < \epsilon_g$}
        \State \textbf{break}
    \EndIf
    \State Construct $G=(V,E)$ with $V=\mathcal{P}$ and $(u,v)\in E$ iff $\mathrm{supp}(\hat{A}_u)\cap\mathrm{supp}(\hat{A}_v)\neq\emptyset$
I    \State $\mathcal{A}_{\mathrm{layer}} \gets \mathrm{MWIS}(G,\boldsymbol{g})$
    \State Add mixer layer $\mathcal{A}_{\mathrm{layer}}$ to the ansatz
    \State $(\boldsymbol{\gamma},\boldsymbol{\beta}) \gets \arg\min_{\boldsymbol{\gamma},\boldsymbol{\beta}} \langle \psi(\boldsymbol{\gamma},\boldsymbol{\beta})|\hat{H}_C|\psi(\boldsymbol{\gamma},\boldsymbol{\beta})\rangle$
    \State $|\psi^{(k)}\rangle \gets |\psi(\boldsymbol{\gamma},\boldsymbol{\beta})\rangle$
    \State $k\gets k+1$
\EndWhile
\State \Return $(\boldsymbol{\gamma}, \boldsymbol{\beta})$, $|\psi\rangle$
\end{algorithmic}
\end{algorithm}

\subsection{Q3SAT-GPT for Circuit Generation}
We detail a QAOA-GPT \cite{tyagin2025qaoa} architecture that generates optimized circuits to bypass the classical optimization overhead. The input to this model is a 3-CNF formula and the output is a tokenized representation of a MosaicADAPT-QAOA circuit. 

\noindent\textbf{Formula Canonicalization.} Aiming to help the model learn similarities between instances, we map the 3-CNF formulas $\mathcal{F}$ to a unique string representation. To achieve this, we apply a deterministic lexicographic sorting to all instances in the dataset. The canonicalization process is defined by two strict ordering rules: (1) \textit{Intra-Clause Literal Ordering:} Within each clause $C_r$, literals $l_{r,j}$ are sorted based on a strict total order $<_L$. The sort key is the variable index $i$. Thus, the literal ordering follows: $x_1 <_L x_2 <_L x_3 <_L \dots <_L x_n$. (2) \textit{Inter-Clause Ordering:} Once the literals within all $m$ clauses are internally sorted, the clauses themselves are sorted lexicographically based on their ordered literals. If two variables have the same index and are of opposite polarity, we consider $x_i <_L \neg x_i$.
An example of this canonicalization process is provided in Appendix \ref{app:canonicalization_example}.

\begin{figure*}[htbp]
    \centering
    \resizebox{\linewidth}{!}{
    \begin{tikzpicture}[
        >=Stealth,
        tok/.style={rectangle, draw, fill=blue!5, rounded corners=2pt, inner sep=4pt, font=\ttfamily\footnotesize, minimum width=22pt, minimum height=22pt, align=center},
        emb/.style={rectangle, draw, fill=orange!10, rounded corners=1pt, minimum width=2.5cm, minimum height=0.6cm, align=center, font=\scriptsize},
        fuse/.style={rectangle, draw, fill=yellow!20, rounded corners=2pt, minimum width=2.5cm, minimum height=0.6cm, font=\scriptsize, align=center},
        final/.style={rectangle, draw, fill=green!10, rounded corners=1pt, minimum width=1.6cm, minimum height=2cm, align=center, font=\scriptsize\bfseries},
        arr/.style={->, thick},
        lbl/.style={font=\footnotesize\itshape, text=gray!80!black}
    ]
    
    \node[font=\large] (formula) at (0, 0) {\textbf{Input Formula:} $(x_1 \lor x_2 \lor x_3) \land (\neg x_2 \lor x_4 \lor \neg x_5)$};
    
    \node[tok, fill=purple!10] (bos) at (-8.5, -1.5) {<bos>};
    \node[tok, right=0.3cm of bos] (t1) {$x_1$};
    \node[tok, right=0.3cm of t1] (t2) {$x_2$};
    \node[tok, right=0.3cm of t2] (t3) {$x_3$};
    \node[tok, fill=red!10, right=0.3cm of t3] (t4) {|};
    \node[tok, right=0.3cm of t4] (t5) {$\neg x_2$};
    \node[tok, right=0.3cm of t5] (t6) {$x_4$};
    \node[tok, right=0.3cm of t6] (t7) {$\neg x_5$};
    \node[tok, fill=red!10, right=0.3cm of t7] (t8) {|};
    \node[tok, fill=purple!10, right=0.3cm of t8] (eof) {<end\_of\_formula>};
    
    \node[tok, fill=purple!10, right=0.3cm of eof] (layer) {<new\_layer\_p>};
    \node[tok, fill=cyan!10, right=0.3cm of layer] (op1)   {$op_1$};
    \node[tok, fill=cyan!10, right=0.3cm of op1]    (b1)  {$\beta_1$};
    \node[tok, fill=cyan!10, right=0.3cm of b1]   (g1)   {$\gamma_1$};
    \node[font=\bfseries, right=0.3cm of g1]      (dots) {\dots};
    
    \draw[thick] ([yshift=0.2cm]bos.north west) -- ([yshift=0.2cm]eof.north east) node[midway, above, font=\footnotesize\bfseries] {Input Formula Sequence};
    \draw[thick] ([yshift=0.2cm]layer.north west) -- ([yshift=0.2cm]g1.north east) node[midway, above, font=\footnotesize\bfseries] {Output Circuit Sequence};
    
    \node[emb]  (te_x1)  at ([yshift=-1.0cm]t1.south) {TokenEmb(x1)};
    \node[emb]  (pe_x1)  at ([yshift=-2.0cm]t1.south) {PosEmb(1)};
    \node[fuse] (mlp_x1) at ([yshift=-3.0cm]t1.south) {$\text{MLP}(\text{FEATHER}(\mathcal{F}_{\text{LCG}}))$};
    
    \node[circle, draw, font=\large\bfseries, inner sep=1pt] (plus_x1) at ([xshift=2.5cm, yshift=-2.0cm]t1.south) {$+$};
    \node[final] (final_x1) at ([xshift=4.2cm, yshift=-2.0cm]t1.south) {$e_{x_1}$\\(Final Emb)};
    
    \draw[arr] (te_x1.east) -| (plus_x1.north);
    \draw[arr] (pe_x1.east) -- (plus_x1.west);
    \draw[arr] (mlp_x1.east) -| (plus_x1.south);
    \draw[arr] (plus_x1.east) -- (final_x1.west);
    
    \draw[arr, dashed, thick, blue] (t1.south) -- (te_x1.north);
    
    \node[emb] (te_lay) at ([yshift=-1.0cm]layer.south) {TokenEmb($\langle$new\_layer\_p$\rangle$)};
    \node[emb] (pe_lay) at ([yshift=-2.0cm]layer.south) {PosEmb(10)};
    \node[fuse] (mlp_lay) at ([yshift=-3.0cm]layer.south) {$\text{MLP}(\text{FEATHER}(\mathcal{F}_{\text{LCG}}))$};
    
    \node[circle, draw, font=\large\bfseries, inner sep=1pt] (plus_lay) at ([xshift=2.0cm, yshift=-2.0cm]layer.south) {$+$};
    \node[final, minimum height=1.6cm] (final_lay) at ([xshift=4.0cm, yshift=-2.0cm]layer.south) {$e_{\text{layer}}$\\(Final Emb)};
    
    \draw[arr] (te_lay.east) -| (plus_lay.north);
    \draw[arr] (pe_lay.east) -- (plus_lay.west);
    \draw[arr] (mlp_lay.east) -| (plus_lay.south);
    \draw[arr] (plus_lay.east) -- (final_lay.west);
    
    \draw[arr, dashed, thick, blue] (layer.south) -- (te_lay.north);
    
    
    \end{tikzpicture}
    }
    \captionsetup{width=\linewidth}
    \caption[Detailed tokenization and embedding depiction.]{Detailed pipeline demonstrating formatting, tokenization, and embeddings. The \texttt{|} delimiter is used to segment clauses, with explicit OR ($\lor$) tokens omitted. 
    }
    \label{fig:tokenization_fusion}
\end{figure*}
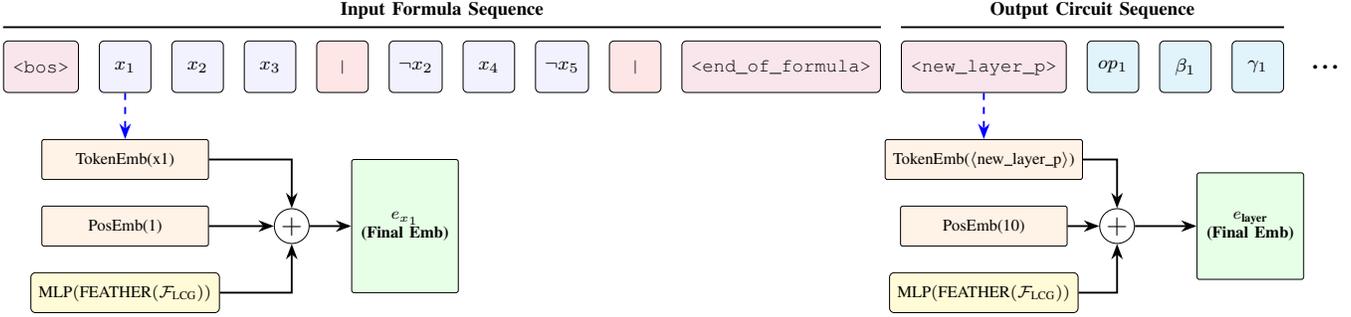

\noindent\textbf{Tokenization.} We adapt the tokenization scheme from QAOA-GPT \cite{tyagin2025qaoa} to 3-CNF formulas and MosaicADAPT-QAOA circuits.
The input sequence contains the canonicalized formula, using tokens $x_i$ and $\neg x_i$ for literals and \texttt{|} as the clause separator, together with the standard $\langle bos\rangle$, $\langle end\_of\_formula\rangle$, and $\langle eos\rangle$ tokens.
To condition the model on global formula structure, we construct a Literal-Clause Graph (LCG) with $2n+m$ nodes, corresponding to the literals and clauses, and connect each literal to the clauses in which it appears.
FEATHER embeddings of this graph are projected through an MLP and added to the token and positional embeddings.
The output sequence follows the QAOA-GPT circuit-token format, except that each MosaicADAPT-QAOA layer contains all disjoint operators selected for that layer, each paired with its corresponding $\beta$ parameter, followed by the layer's $\gamma$ parameter.
Thus, a layer is represented as $\langle new\_layer\_p\rangle,\text{op}_1,\beta_1,\ldots,\text{op}_r,\beta_r,\gamma$.
See Figure \ref{fig:tokenization_fusion} for a visual representation.

\noindent\textbf{Loss Function.} To ensure that the model is trained only on predicting valid and optimized circuit sequences, we modify the Categorical Cross Entropy loss function used in QAOA-GPT to add the loss only for the output tokens as:
    \begin{equation}\label{eq:loss_function}
    L = -\sum_{t} \mathbbm{1}_{\{y_t \neq 0\}} \log(p(y_t | X_{\leq t})),
    \end{equation}
    where $\mathbbm{1}_{\{y_t \neq 0\}}$ is an indicator function that masks the loss for parameters associated with the input formula and padding tokens (target index 0). In contrast to the original QAOA-GPT model, we do not add a sliding window to extract subsequences, but rather use a single context window of 1024 tokens, ensuring that all the sequences in the dataset fit in the context window, including the input formula, the output circuit and the separator tokens. The tokens after the $\langle eos \rangle$ token are padded with $\langle pad \rangle$ tokens, and the loss is masked for the $\langle pad \rangle$ tokens. 
    
    \noindent\textbf{Inference and Circuit Decoding.} 
    During inference, Q3SAT-GPT generates optimized quantum circuits in a single forward pass. Given a tokenized formula, the model predicts a sequence of tokens autoregressively. At each step, a token is sampled from the output distribution using a temperature $T$ to control diversity. The predicted tokens are mapped back to quantum gates (operators) and their corresponding variational parameters ($\boldsymbol\gamma, \boldsymbol\beta$). We consider a circuit sequence as the tokens between the $\langle end\_of\_formula\rangle$ and the $\langle eos\rangle$ tokens.

\section{Experimental Setup}

\subsection{Evaluation Metrics}
\noindent\textbf{Approximation Ratio.} The quality of the results is measured in terms of Approximation Ratio (AR), where the AR of a solution for a formula is defined as
$$\text{AR} = \frac{\text{Number of clauses satisfied by a candidate method}}{\text{Maximum number of clauses that can be satisfied}} $$
The maximum number of clauses that are satisfiable was computed using Gurobi \cite{gurobi}.

\noindent\textbf{Error Rate.} The structural validity of the GPT generated circuits is measured using Error Rate (ER). For each problem instance, the GPT model can generate different circuits based on its temperature parameter. We sample 5 circuits from the model for each formula. For generated-circuit results, AR is computed over valid generated circuits, while ER is reported at both the formula and circuit levels. The Formula Error Rate denotes the percentage of formulas in which no sampled circuit was valid for the formula. The Circuit Error Rate denotes the total percentage of sampled circuits that were invalid.

\subsection{Setup for MosaicADAPT-QAOA}
\noindent\textbf{Baselines for Adaptive Circuit Construction.}
To evaluate the efficacy of MosaicADAPT-QAOA, we compare it against two primary adaptive baselines: (1) \textit{ADAPT-QAOA}, which selects a single operator per layer based on the highest individual gradient, and (2) \textit{TETRIS-QAOA}, which is a modification of ADAPT-QAOA and employs a greedy selection of multiple disjoint operators. All three methods use the same operator pool defined in Section \ref{sec:operator_pool} and the same hyperparameters, which are detailed in Appendix \ref{stopping_criteria}. 
The algorithm's termination is primarily dictated by two dominant criteria: the convergence of all operator gradients below $10^{-6}$ or the reaching of the 20-layer maximum adaptation limit.

\noindent\textbf{Benchmark Test Set.} We generated a benchmark set of 100 Max-E3-SAT instances. Each instance contains 10 variables and is satisfiable. 
50 of the 100 instances are generated using the Uniform Random generation process, and the other 50 instances are from the Balanced distribution \cite{spence2017balanced} generated using the \texttt{satqubolib} Python library~\cite{satqubolib}.

For balanced instances, the number of clauses $m$ for an instance with $n$ (here $n=10$) variables is drawn from a randomized ratio around the phase transition value $\alpha_0 = 3.6$:
\begin{equation}
    \alpha \;=\; \alpha_0 \,(1 + \delta), \qquad \delta \sim \mathrm{Uniform}(-0.2,\; 0.2),
    \label{eq:ratio}
\end{equation}
so that $\alpha \in [2.88,\; 4.32]$ (i.e.\ $\alpha_0 \pm 20\%$). For the uniform random instances, the number of clauses $m$ for an instance with $n$ variables is drawn from a uniform distribution between $\alpha = 4.26 \pm 20\%$.


\subsection{Setup for Q3SAT-GPT}
\noindent\textbf{Dataset Construction.}
We conduct our Q3SAT-GPT training on two problem sizes: 10- and 12-variable SAT instances.
For each problem size, we generate a dataset containing 50000 Max-E3-SAT instances.
It consists of 25K uniform random instances and 25K balanced instances.
For each type, half of the instances are satisfiable and half are unsatisfiable.
We ensure that each instance is unique after canonicalization.
The dataset finally is split into train (80\%), validation (10\%), and test (10\%) partitions stratified by satisfiability and formula generation method.

\noindent\textbf{Model Variants.} We train a total of 4 models each, for 10-variable and 12-variable Max-E3-SAT: 
\begin{enumerate}
    \item Without LCG-graph embeddings: A pure transformer model that generates circuit parameters conditioned exclusively on the literal tokenized 3-CNF formula.
    \item With FEATHER embeddings on LCG-graph embeddings: The FEATHER embeddings are projected to the input embedding layer through a multi-layer perceptron, aiming to provide global context about the formula. These embeddings are added to the token and positional embeddings.
\end{enumerate}
For each of the above variant, we train two models based on the initial gamma $\gamma_0$ hyperparameter:
\begin{enumerate}
    \item $\gamma_0=0.5$: An initial gamma of 0.5 is used for producing all 50K optimized circuits using MosaicADAPT-QAOA.
    \item $\gamma_0 = \mathrm{argmax}_{\gamma \in \{0.01, 0.1, 0.5\}}\mathrm{AR}(\gamma)$: A grid search over different initial gamma values is performed for all 50K instances. The circuit yielding the highest AR is selected, with ties broken by fewer layers.
\end{enumerate}

Additional training implementation details, including hardware, architecture, context window, dropout, and validation-loss cadence, are provided in Appendix \ref{app:q3sat_gpt_training_details}.

\noindent\textbf{Model Selection.} We perform model selection on the AR metric using a fixed validation set.
50 validation set formulas are randomly sampled from the validation partition at startup with a fixed seed, held fixed for the entire training run. 
For each instance, Q3SAT-GPT generates 5 circuits autoregressively.
The generated circuits are then evaluated 
on the AR and ER metrics. 
The model with the highest AR and lowest ER is selected for deployment on the test set.

\section{Results}
\subsection{MosaicADAPT-QAOA}
\begin{figure}[htbp]
    \centering
    \begin{subfigure}{0.5\textwidth}
        \centering
        \includegraphics[width=\linewidth]{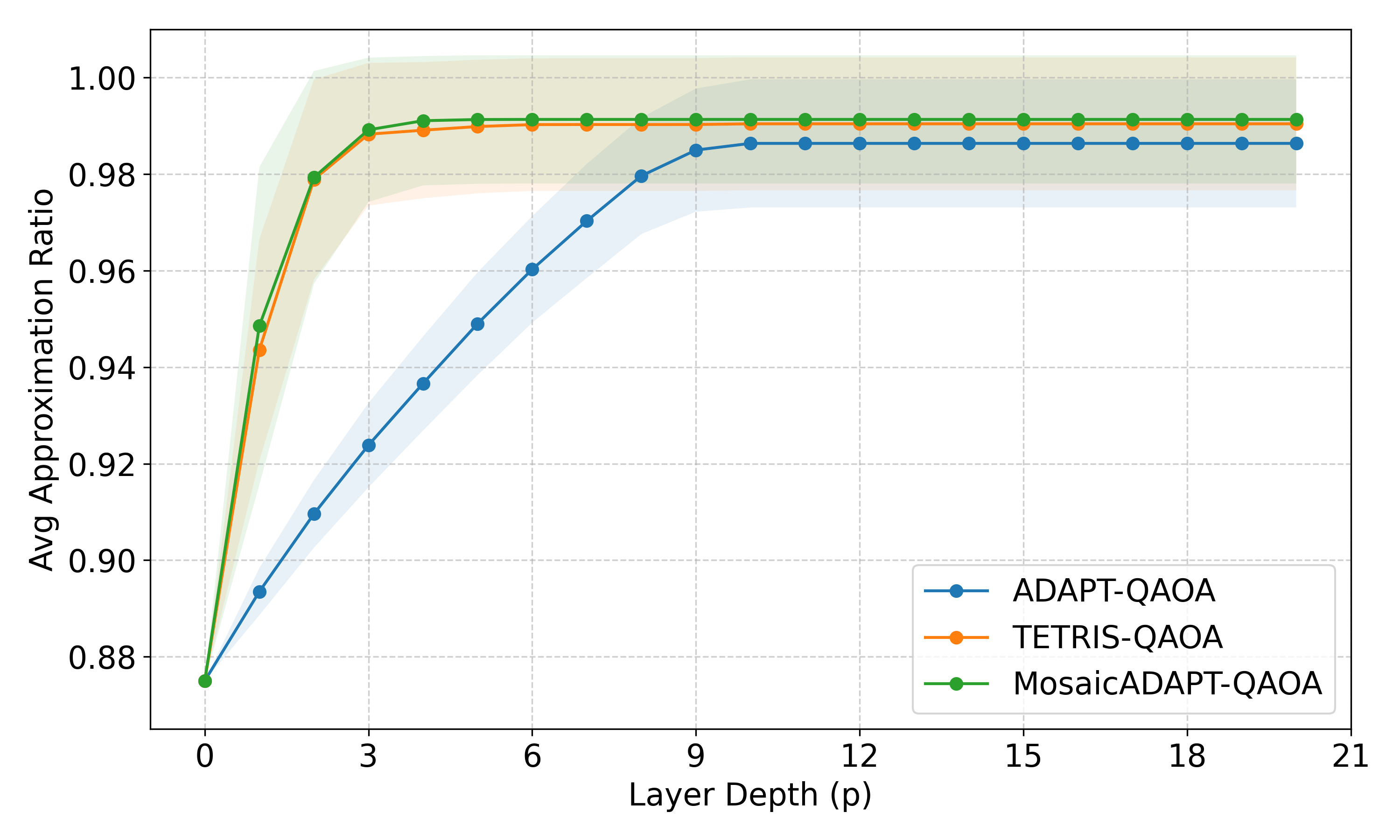}
        \caption{$\gamma_0 = 0.01$}
        \label{fig:conv_low}
    \end{subfigure}
    \hfill
    \begin{subfigure}{0.5\textwidth}
        \centering
        \includegraphics[width=\linewidth]{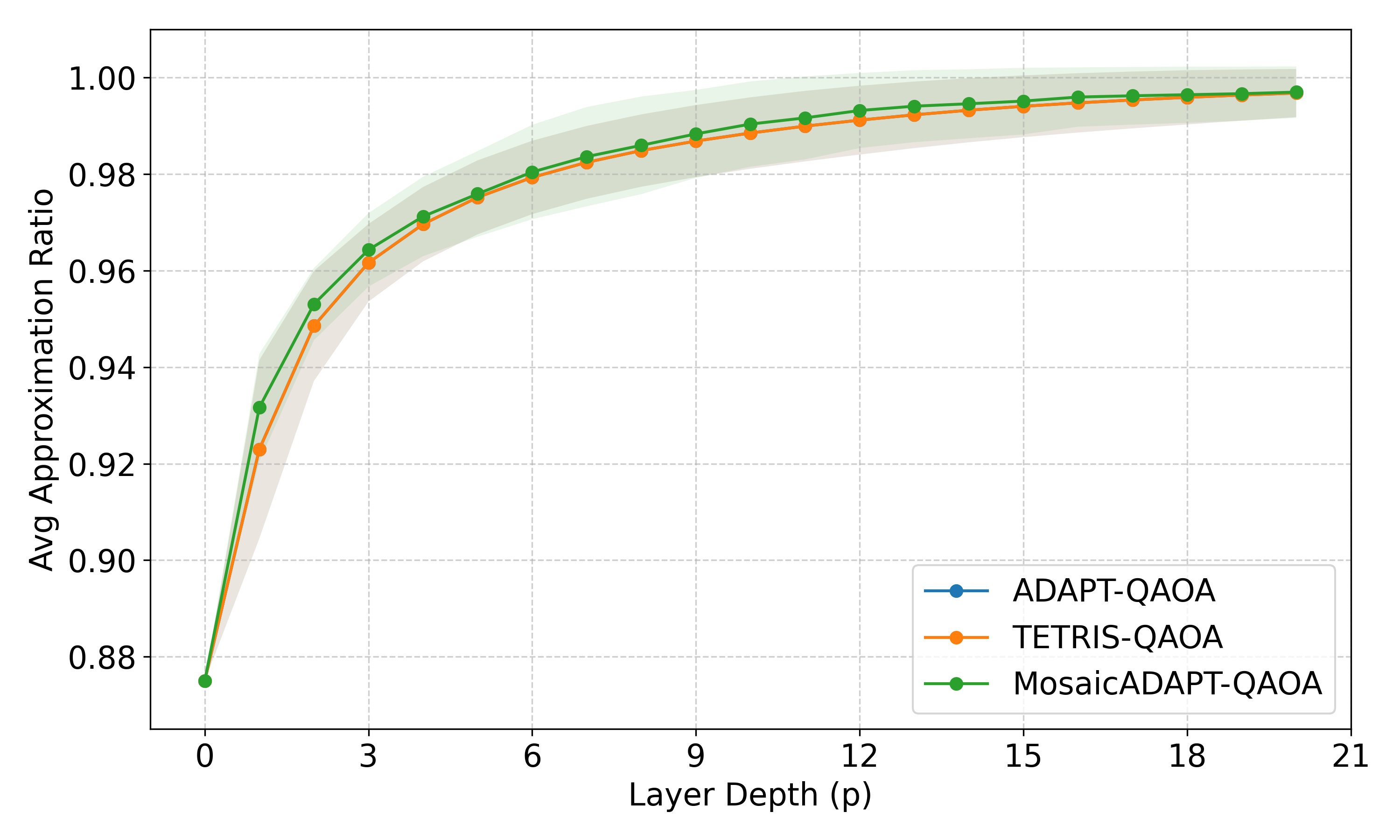}
        \caption{$\gamma_0 = 0.5$}
        \label{fig:conv_high}
    \end{subfigure}
    \caption[Comparison of the energy convergence across adaptations for a low ($\gamma_0 = 0.01$) and high ($\gamma_0 = 0.5$) initial gamma value.]{Comparison of the energy convergence across adaptations. Note that in Figure (b), plots of TETRIS-QAOA and ADAPT-QAOA overlap due to the same operators being chosen by both the methods (see Figure \ref{fig:op_evol_comparison}).}
    \label{fig:conv_comparison}
\end{figure}

We compare the performance of the three methods (ADAPT-QAOA, TETRIS-QAOA, MosaicADAPT-QAOA) at a low initial gamma value, i.e., $\gamma_0 = 0.01$, and a relatively higher initial gamma value, i.e., $\gamma_0 = 0.5$, to understand the behavior of the algorithms. Overall, we observe that MosaicADAPT-QAOA performs better in terms of clause satisfaction rate and circuit depth than TETRIS-QAOA and ADAPT-QAOA. 
Table \ref{tab:variant_comparison} shows that the mean and median clause satisfaction rates for MosaicADAPT-QAOA are higher than those of TETRIS-QAOA and ADAPT-QAOA. At any particular layer, the AR of MosaicADAPT-QAOA is higher on average (Figure \ref{fig:conv_comparison}). 



\begin{table}[ht]
\centering
\small
\setlength{\tabcolsep}{5pt}
\caption[Comparison of clause satisfaction rates for ADAPT-QAOA, TETRIS-QAOA, and MosaicADAPT-QAOA]{Comparison of Performance on Different ADAPT variants. The Median number of layers to reach 99.9\% AR are shown only for the instances that reached this target within 20 layers.}
\label{tab:variant_comparison}
\begin{tabular}{lcccc}
\toprule
Method & \makecell{Mean \\AR\\(\%)} & \makecell{Median \\AR\\(\%)} & \makecell{Median \\ Layers \\ to 99.9\% \\AR} & \makecell{Fraction\\ of \\Stuck\\ Instances} \\
\midrule
\multicolumn{5}{c}{\textit{$\gamma_0 = 0.01$}} \\
\addlinespace[0.5ex]
ADAPT-QAOA & 98.64 & 98.62 & 9 & 13/100 \\
TETRIS-QAOA & 99.05 & \textbf{100.00} & \textbf{2} & 23/100 \\
MosaicADAPT-QAOA & \textbf{99.13} & \textbf{100.00} & \textbf{2} & 27/100 \\
\midrule
\multicolumn{5}{c}{\textit{$\gamma_0 = 0.5$}} \\
\addlinespace[0.5ex]
ADAPT-QAOA & 99.68 & 99.88 & 16 & 0/100 \\
TETRIS-QAOA & 99.68 & 99.88 & 16 & 0/100 \\
MosaicADAPT-QAOA & \textbf{99.71} & \textbf{100.00} & \textbf{12} & 0/100 \\
\bottomrule
\end{tabular}
\end{table}

\begin{figure}[htbp]
    \centering
    \begin{subfigure}{0.5\textwidth}
        \centering
        \includegraphics[width=\linewidth]{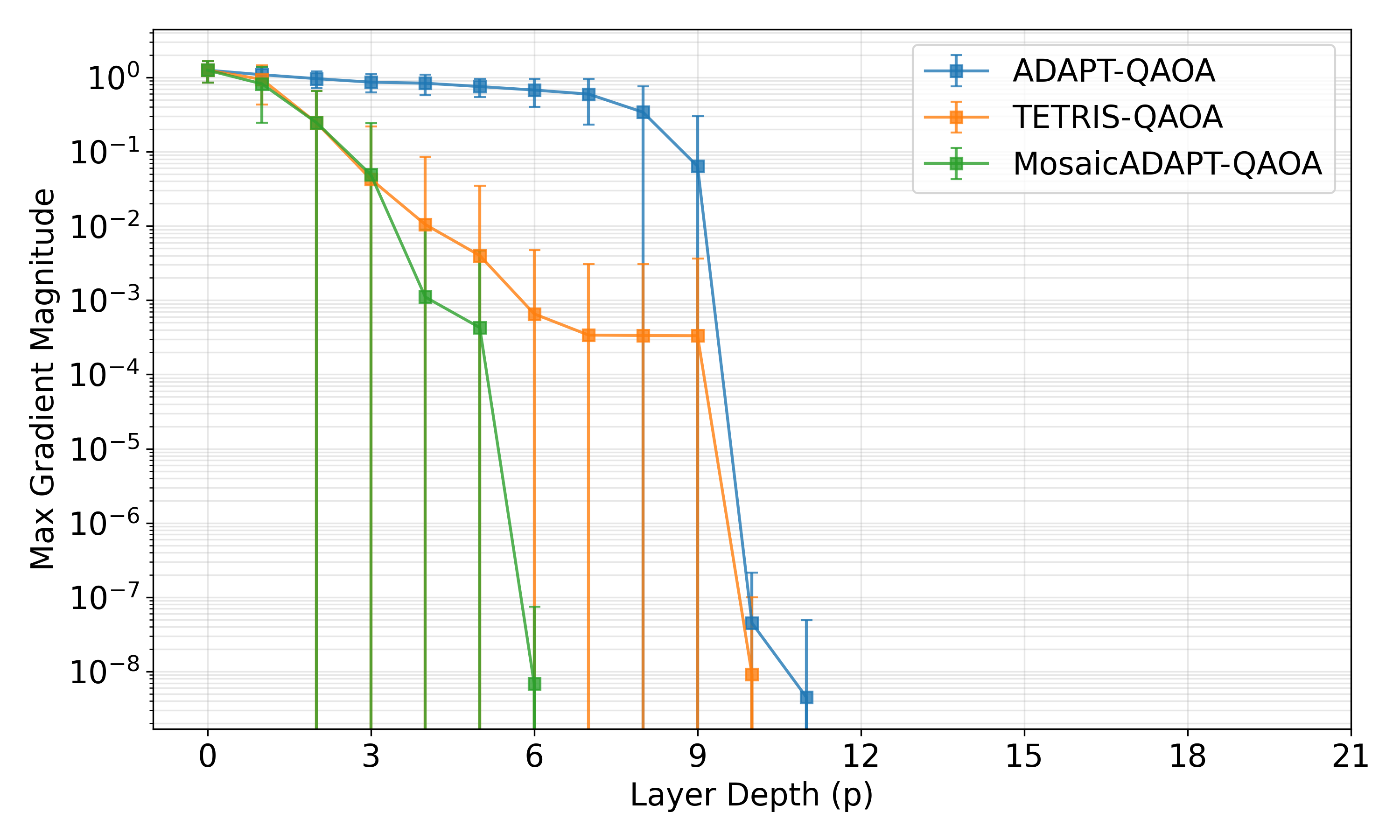}
        \caption{$\gamma_0 = 0.01$}
        \label{fig:max_grad_low}
    \end{subfigure}
    \hfill
    \begin{subfigure}{0.5\textwidth}
        \centering
        \includegraphics[width=\linewidth]{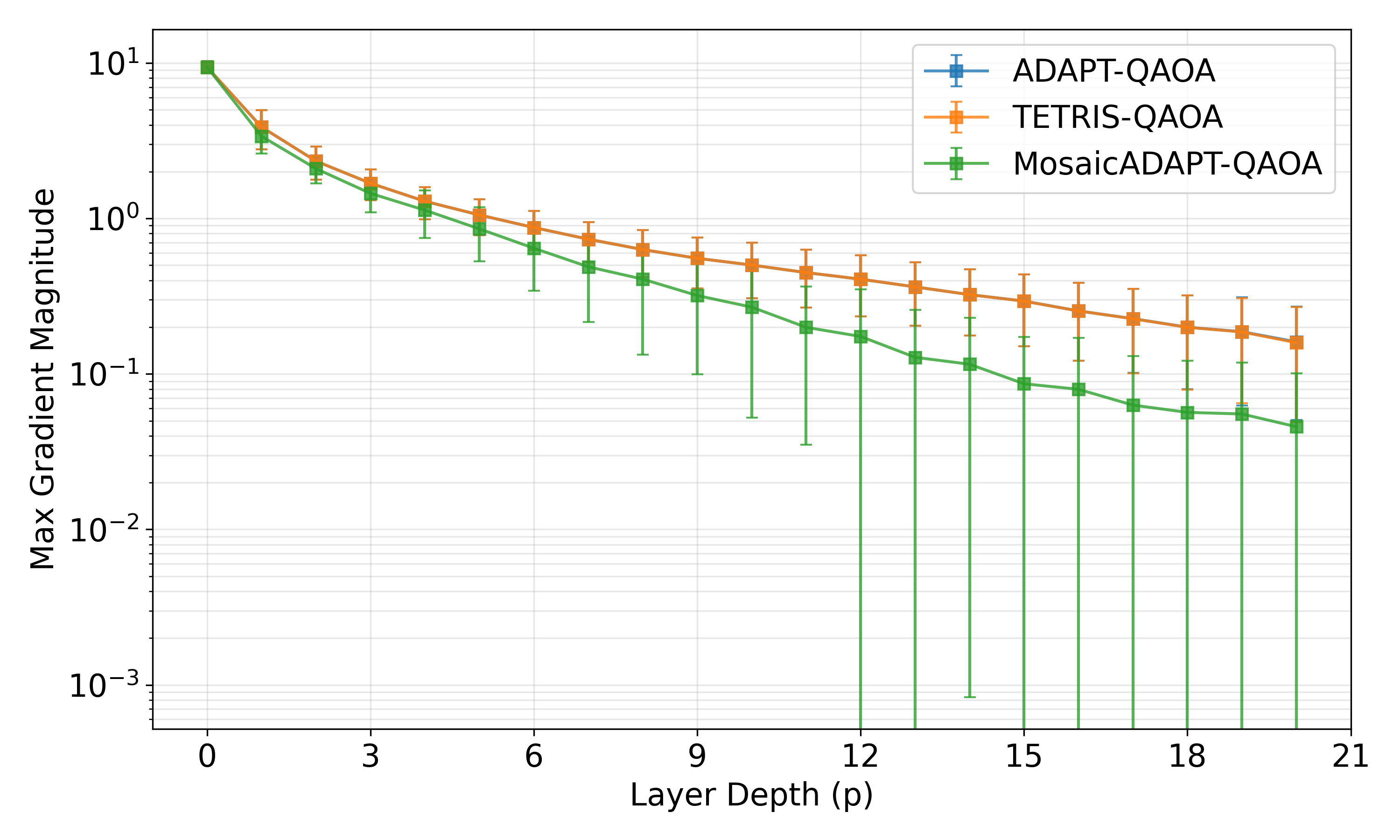}
        \caption{$\gamma_0 = 0.5$}
        \label{fig:max_grad_high}
    \end{subfigure}
    \caption{Comparison of the maximum gradient norm across all operators in the mixer pool. Note that in Figure (b), plots of TETRIS-QAOA and ADAPT-QAOA overlap due to the same operators having the maximum gradient for the most part (see Figure \ref{fig:op_evol_comparison}).}
    \label{fig:max_grad_comparison}
\end{figure}

\noindent\textbf{Sensitivity to Initial Gamma.}
For low initial gamma values, we observe that all 3 methods are prone to early stopping due to a sudden drop in gradients below the accepted threshold (Figure \ref{fig:max_grad_comparison}).
We define an instance as ``stuck'' if in 1000 shot sampling, the energies of all final bitstrings are strictly greater than the ground state energy. Table \ref{tab:variant_comparison} shows that $\gamma_0 = 0.01$ led to more frequent such cases where instances were stuck due to early stopping. Importantly, in cases where the tendency of getting stuck is less ($\gamma_0=0.5$), MosaicADAPT-QAOA required a median of 4 fewer layers than the other variants.


\subsection{Q3SAT-GPT Results}
\noindent\textbf{Performance on Test Instances.} Table \ref{tab:q3sat_gpt} presents the results on the Approximation Ratio and Error Rate metrics. We observe that, in general, the circuits generated with FEATHER embeddings have a lower error rate, though the difference in performance in terms of AR in inconclusive. 
The AR achieved on balanced instances is consistently higher than that on random instances, suggesting that the model is able to differentiate between the types of instances and that there is a structure to the resultant circuits for the balanced instances. In addition, the model performed better when trained on a dataset optimized with a single $\gamma_0 = 0.5$ rather than a grid search. \begin{table*}[h]
\centering
\caption{Comparison of Model Training Methodologies}
\label{tab:q3sat_gpt}
\small
\setlength{\tabcolsep}{4pt} 
\begin{tabular}{llc cc cc cc cc c} 
\toprule
& & & \multicolumn{2}{c}{\textbf{Formula ER (\%)}} & \multicolumn{2}{c}{\textbf{Circuit ER (\%)}} & \multicolumn{2}{c}{\textbf{Best AR (\%)}} & \multicolumn{2}{c}{\textbf{Avg AR (\%)}} & \multirow{2}{*}{\shortstack{\textbf{MosaicADAPT}\\\textbf{QAOA AR (\%)}}} \\
\cmidrule(lr){4-5} \cmidrule(lr){6-7} \cmidrule(lr){8-9} \cmidrule(lr){10-11}
\textbf{N} & \textbf{Selected Instance} & \textbf{Distribution} & \textbf{F} & \textbf{W/O} & \textbf{F} & \textbf{W/O} & \textbf{F} & \textbf{W/O} & \textbf{F} & \textbf{W/O} & \\
\midrule

\multirow{4}{*}{10} & \multirow{2}{*}{\shortstack[c]{Grid Search}} 
 & Balanced  & 0.00 & 0.00 & 5.00 & 4.14 & 98.41 & 98.52 & 97.02 & 97.36 & 99.83 \\
 & & Random  & 0.00 & 0.00 & 0.60 & 0.12 & 95.17 & 96.03 & 93.85 & 94.42 & 99.96 \\
\cmidrule(lr){2-12}
 & \multirow{2}{*}{\shortstack[c]{$\gamma_0=0.5$}} 
 & Balanced & 0.00  & 0.00  & 0.05 & 1.16 & 98.88 & 99.00  & 98.79 & 98.83 & 99.42 \\
 & & Random    & 0.76 & 6.56 & 9.82 & 16.10 & 97.83 & 92.60  & 96.39 & 91.37 & 99.84 \\

\midrule

\multirow{4}{*}{12} & \multirow{2}{*}{\shortstack[c]{Grid Search}} 
 & Balanced  & 0.00 & 0.00 & 0.53 & 0.45 & 97.04 & 98.55 & 94.18 & 98.21 & 99.77 \\
 & & Random  & 0.00 & 0.00 & 0.61 & 1.66 & 93.98 & 95.34 & 92.41 & 94.18 & 99.96 \\
\cmidrule(lr){2-12}
 & \multirow{2}{*}{\shortstack[c]{$\gamma_0=0.5$}} 
 & Balanced & 0.00 & 0.00 & 0.01 & 0.01 & 97.28 & 98.62 & 96.94 & 98.51 & 99.34 \\
 & & Random & 0.00 & 0.00 & 1.30 & 5.30  & 96.53 & 97.81 & 94.89 & 95.90  & 99.85 \\

\bottomrule
\multicolumn{12}{l}{\footnotesize \textbf{N}: Number of variables in MAX-E3-SAT, \textbf{ER}: Error Rate, \textbf{AR}: Approximation Ratio , \textbf{F}: With FEATHER, \textbf{W/O}: Without FEATHER.} \\
\multicolumn{12}{l}{\footnotesize AR values are calculated only for valid circuits and are expressed as a percentage}
\end{tabular}
\end{table*}

\noindent\textbf{Inference Time Comparison.} Table \ref{tab:runtime_performance} shows the amortized inference time for Q3SAT-GPT predicted circuits and compares it against the time required to find a single MosaicADAPT-QAOA circuit on a single-threaded CPU. Q3SAT-GPT performance is reported as amortized inference time, calculated by dividing the total batch execution time by the batch size ($B=128$).
\begin{table}
\centering
\caption{Runtime Performance Comparison.}
\label{tab:runtime_performance}
\small
\begin{tabular}{l rr}
\toprule
\textbf{Method} & \textbf{N=10 (s)} & \textbf{N=12 (s)} \\
\midrule
\textit{Amortized Inference (GPU)} & & \\
Q3SAT-GPT & 0.16 & 0.45 \\ 
\midrule
\textit{Sequential Optimization (CPU)} & & \\
MosaicADAPT-QAOA & 88.21 & 503.72 \\ 
\bottomrule
\end{tabular}
\end{table}

\section{Discussion}
\subsection{Circuit Structure Sensitivity to Initial Gamma}
\begin{figure*}[htbp]
    \centering
    \begin{subfigure}{\textwidth}
        \centering
        \includegraphics[width=\linewidth]{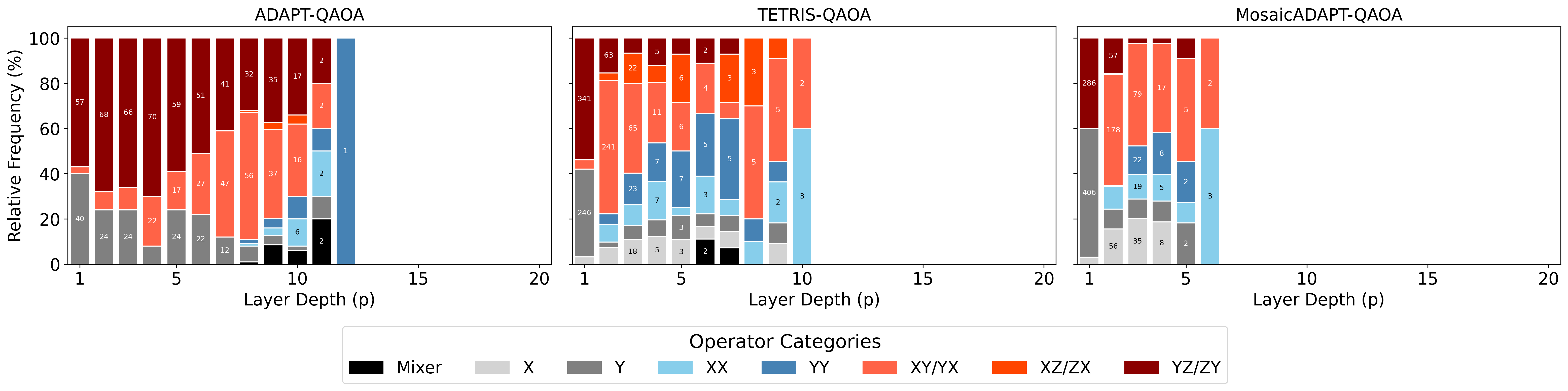}
        \caption{$\gamma_0 = 0.01$}
        \label{fig:op_evol_low}
    \end{subfigure}
    \\[2ex]
    \begin{subfigure}{\textwidth}
        \centering
        \includegraphics[width=\linewidth]{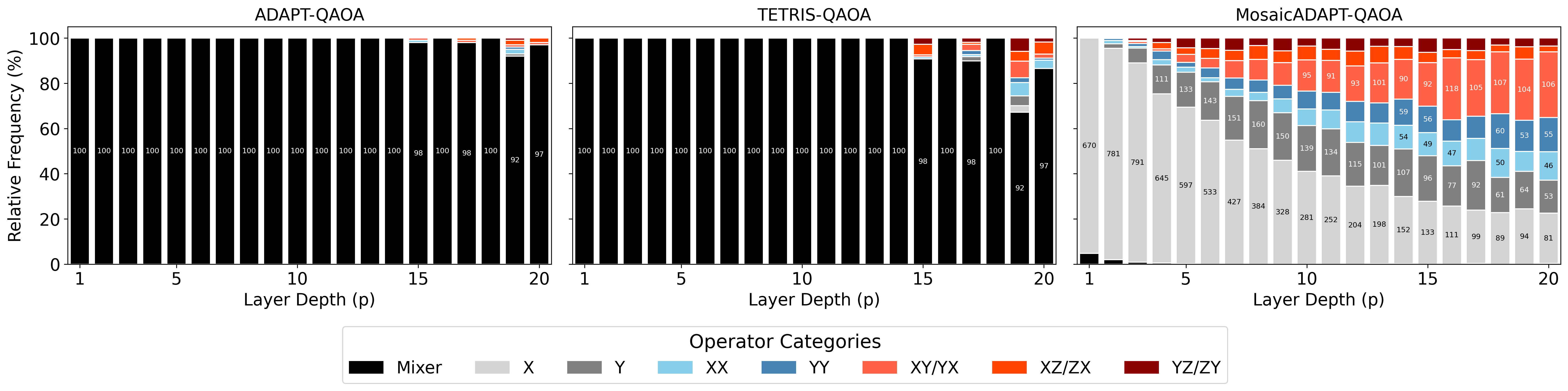}
        \caption{$\gamma_0 = 0.5$}
        \label{fig:op_evol_high}
    \end{subfigure}
    \caption[Evolution of the operator selection across adaptations for a low ($\gamma_0 = 0.01$) and high ($\gamma_0 = 0.5$) initial gamma value.]{Evolution of the operator selection across adaptations. Note that the count inside a bar is that of the total number of operators selected in that layer across all instances.}
    \label{fig:op_evol_comparison}
\end{figure*}

We often observe that for low initial values of gamma ($\gamma_0 < 0.1$), the classical optimizer pushes the gamma values to be close to 0. It also pushes the beta values to $\pm \frac{\pi}{4}$. This is in line with recent studies that observe the same effect, and that ADAPT-QAOA is prone to get stuck in an excited state \cite{sridhar2023adapt}. Additionally, at low initial $\gamma$ values, all three methods mostly choose Y-type operators or YZ-type operators (Figure \ref{fig:op_evol_comparison}). With the corresponding $\beta$ value of $\pm \frac{\pi}{4}$, these operators make a qubit go from a uniform superposition of $|0\rangle$ and $|1\rangle$ to either $|0\rangle$ or $|1\rangle$. Since the $\gamma$ values are pushed to 0, the behavior of the optimized circuit does not seem to leverage the phase changes induced by the cost Hamiltonian in QAOA. Since the circuits optimized on low initial $\gamma$ values had final $\beta$ values close to $\pm \frac{\pi}{4}$ and final gamma values close to 0, Q3SAT-GPT does not seem to yet find out which qubit in a uniform superposition should be pushed to a computational basis state. This might explain that the  performance of Q3SAT-GPT with a grid-search could be improved. 
On the other hand, at higher initial $\gamma_0$ values, we observe that ADAPT-QAOA and TETRIS-QAOA mostly choose the standard QAOA mixer in the initial mixer layers. 
However, MosaicADAPT-QAOA chooses the single-qubit X-operators and essentially moves towards multi-angle QAOA in the initial layers. As the algorithm 
progresses, it makes use of the other operators in the pool in the subsequent layers. This allows MosaicADAPT-QAOA to perform better than ADAPT-QAOA and TETRIS-QAOA at $\gamma_0=0.5$.

\subsection{Depth vs. Parameter Tradeoff}
We distinguish between adaptive layers, selected mixer operators, and variational parameters. MosaicADAPT-QAOA reduces the number of adaptive layers by selecting denser compatible operator sets, but this can increase the number of trainable parameters, especially in the $\gamma_0$ = 0.5 regime where ADAPT-QAOA and TETRIS-QAOA often select the standard QAOA mixer in the initial layers.
While ADAPT-QAOA and TETRIS-QAOA required a median number of parameters of 17 and 17.5 respectively, the MosaicADAPT-QAOA method required a median of 97 parameters to reach the 99.9\% AR. 
This motivates the use of generative AI based approaches. If the GPT model is able to learn the parameters, the massive parameter optimization overhead is mitigated, while still learning from a rich dataset.


\subsection{Q3SAT-GPT Predicted Circuit Structure}

\begin{figure}[htbp]
     \centering
     \begin{subfigure}[b]{0.98\columnwidth}
         \centering
         \includegraphics[width=\textwidth]{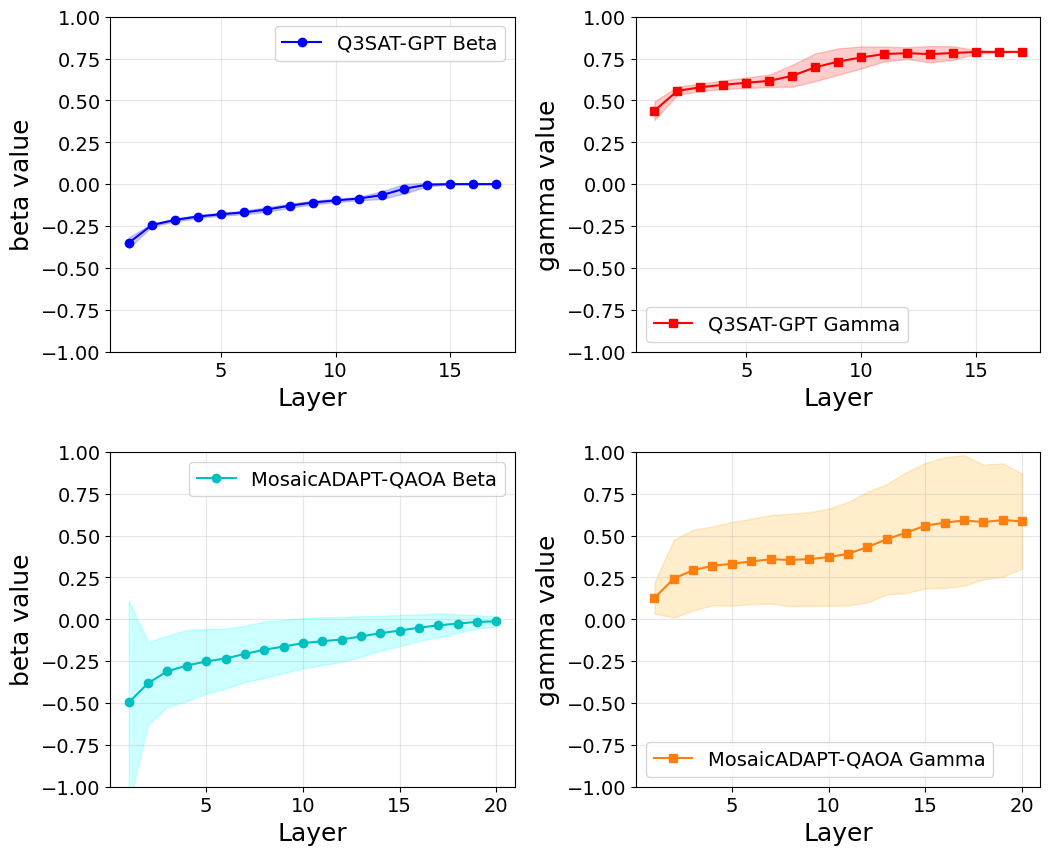}
         \caption{Balanced instances.}
         \label{fig:left_image}
     \end{subfigure}
     \\ 
     \begin{subfigure}[b]{0.98\columnwidth}
         \centering
         \includegraphics[width=\textwidth]{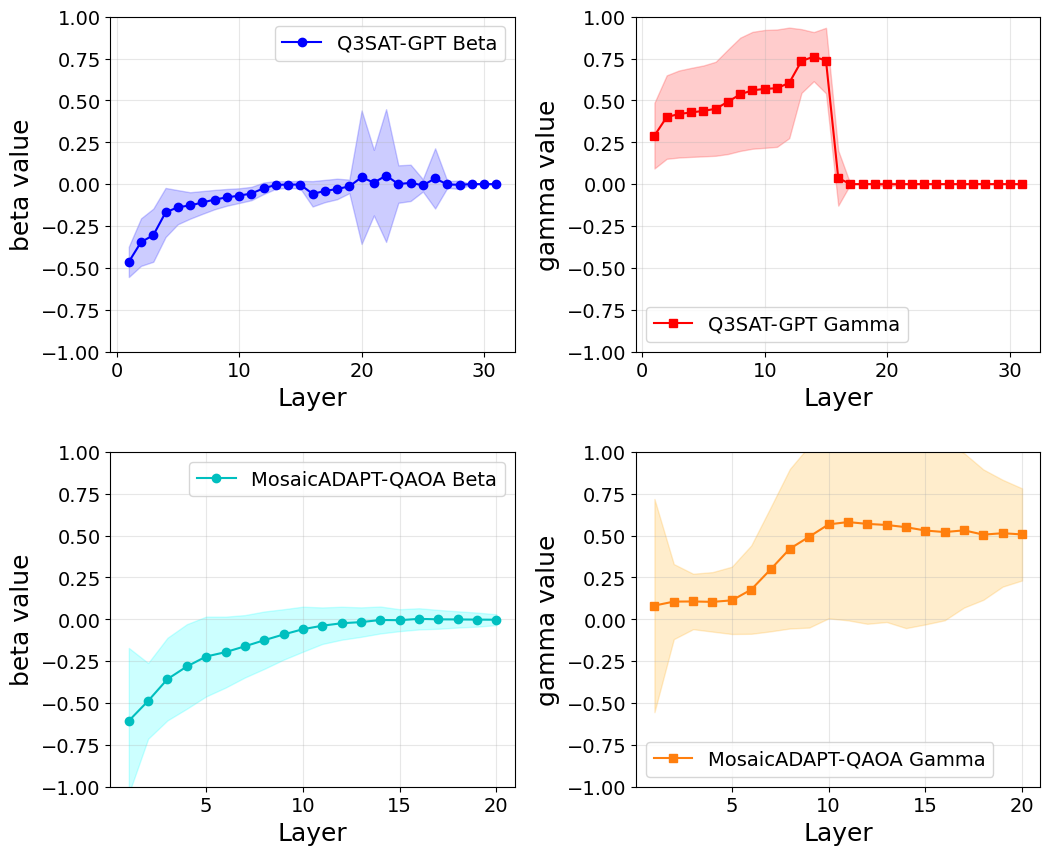}
         \caption{Random instances.}
         \label{fig:right_image}
     \end{subfigure}
     
     \caption{Comparison of Q3SAT-GPT predicted parameters against MosaicADAPT-QAOA optimized parameters on the test set. Solid lines denote the mean parameter value across samples, and shaded regions denote $\pm 1$ standard deviation. This plot is for the configuration containing 12 qubits, with $\gamma_0 = 0.5$, and utilizes FEATHER embeddings. Since there are multiple beta values at a particular layer for MosaicADAPT-QAOA, all beta values are averaged at that layer.}
     \label{fig:side_by_side_comparison}
\end{figure}

Figure \ref{fig:side_by_side_comparison} compares the variance of $\gamma$ and $\beta$ values for circuits generated by Q3SAT-GPT and circuits optimized directly by MosaicADAPT-QAOA.
For balanced instances, the model often predicts highly concentrated parameter values, yet these circuits still reach a 98.51\% approximation ratio.
For random instances, the generated circuits show more variability, and in a few cases the model does not predict the $\langle eos\rangle$ token by the 20th layer.
Another issue is that the model also does not consistently reproduce the late-layer operators selected by MosaicADAPT-QAOA.
This does not always harm solution quality because the corresponding late-layer $\beta$ values are often near zero.
However, generating higher-quality circuits will likely require better recovery of the operators selected in the final adaptive layers.

\subsection{Lessons Learned and Future Directions}
The generated circuits suggest that the structure of the MosaicADAPT-QAOA training data strongly shapes the behavior of Q3SAT-GPT.
In the training set, the first few layers are often dominated by single-qubit Pauli-X operators, while operators that appear in later adaptive layers are much less frequent.
As a result, a model trained with categorical cross-entropy can overrepresent early-layer patterns when generating later layers.
This suggests that future models should treat layer position and operator diversity more explicitly, for example through layer-aware losses, reweighting of underrepresented late-layer operators, or targeted augmentation of training sequences.

Syntactic validity and circuit quality should also be separated more clearly.
The current model can occasionally generate structurally invalid circuits, especially at larger depths where the output sequence is longer and the set of valid next tokens is more constrained.
Constrained decoding would allow the model to sample only structurally valid operators, parameters, and layer delimiters at each generation step.
These constraints would not guarantee high approximation ratio by themselves, but they would remove avoidable structural errors and let model capacity focus on circuit quality rather than circuit grammar.
Repetition penalties or diversity-promoting decoding rules may also help prevent the model from repeatedly selecting the same dominant operator types in later layers.

The performance gap between balanced and random instances also suggests that richer instance-level information remains important.
Q3SAT-GPT appears to learn circuit patterns more easily when formulas have more regular structure.
SAT-specific neural architectures such as NeuroSAT \cite{neurosat} or SAT-GATv2 \cite{satgatv2} could provide stronger inductive bias than the current literal-clause graph embeddings alone, as related graph-based embeddings have helped in QAOA parameter learning \cite{falla2024graph}.
Such embeddings may help the model distinguish which adaptive operators are useful for a particular formula, rather than relying primarily on global dataset-level circuit patterns.

Finally, Q3SAT-GPT should be viewed as a first step towards closed-loop generative quantum circuit discovery.
The present model learns from offline MosaicADAPT-QAOA trajectories and generates circuits without classical variational optimization.
A natural next step is to add feedback from circuit execution, either through simulation or quantum hardware, so that generated circuits can be reinforced according to measured solution quality.
Such a feedback loop could combine the speed of autoregressive generation with quality signals from actual circuit performance. Finding a balance between them that would keep the method scalable is a challenging future direction. 


\section*{Acknowledgments}

This research was supported in part through the use of DARWIN
computing system funded by NSF Grant \#1919839. 
K. Shirali was supported by the U.S. Department of Energy, Office of Science, National Quantum Information Science Research Centers, Co-design Center for Quantum Advantage (C2QA) under Contract Number DE-SC0012704. This research was supported in part by NSF Grant \#2427042.

\clearpage
\appendices

\section{Technical details}
\subsection{Stopping Criteria}\label{stopping_criteria}
ADAPT-QAOA, in particular the implementation in \href{https://github.com/KarunyaShirali/ADAPT.jl.git}{ADAPT.jl}, uses additional hyperparameters to control the stopping criteria and convergence of the algorithm. These hyperparameters are detailed in Table \ref{tab:hyperparams_results}, along with the description of the hyperparameter, and the corresponding value used in our study.

\subsection{Canonicalization Example}\label{app:canonicalization_example}
The following example illustrates the intra-clause and inter-clause sorting rules used to canonicalize each 3-CNF formula before tokenization. 
Given an unsorted formula $\mathcal{F}$:
$$(x_2 \lor \neg x_4 \lor x_3) \land (x_1 \lor x_3 \lor x_2) \land (x_1 \lor x_2 \lor \neg x_4),$$
the intra-clause sorting first reorders the literals within each clause:
$$(x_2 \lor x_3 \lor \neg x_4) \land (x_1 \lor x_2 \lor x_3) \land (x_1 \lor x_2 \lor \neg x_4).$$
Then, the inter-clause sorting orders the clauses lexicographically to produce the final canonical string fed to the model:
$$(x_1 \lor x_2 \lor x_3) \land (x_1 \lor x_2 \lor \neg x_4) \land (x_2 \lor x_3 \lor \neg x_4).$$

\subsection{Q3SAT-GPT Training Details}\label{app:q3sat_gpt_training_details}
The training is performed on a single NVIDIA A100 GPU on the University of Delaware Caviness cluster.
The transformer architecture is based on nanoGPT and consists of 6 attention layers with 6 attention heads and 384-dimensional embeddings.
The context window is set to 1024 tokens and the dropout rate is set to 0.2.
Validation loss is estimated every 100 iterations by randomly sampling 2 batches (1 batch contains 512 sequences) from the validation split.

\begin{table*}[h]
\centering
\small
\begin{tabular}{llp{8cm}}
\toprule
\textbf{Parameter} & \textbf{Value} & \textbf{Description} \\
\midrule
\texttt{layer\_stopper\_max} & $20$ & Maximum Adaptations (Number of layers) permitted in the circuit \\
\texttt{floor\_stopper\_energy} & $-\infty$ & Floor energy ($E_{\mathrm{floor}}$): a target energy which we would be satisfied with. We set it to $-\infty$ to understand the convergence of the algorithm. \\
\texttt{floor\_stopper\_threshold} & $0.1$ & ADAPT stops if the expected energy is within threshold of the Floor energy, $E \leq E_{\mathrm{floor}} + 0.1$ (Not used in current study) \\
\texttt{optimizer\_tolerance} & $10^{-6}$ & Gradient threshold below which BFGS optimizer reaches convergence criterion when optimizing the parameters \\
\texttt{optimizer\_max\_iterations} & $1000$ & Maximum optimiser iterations for optimizing the parameters for a fixed number of layers after an adaptation is performed \\
\texttt{slow\_stopper\_threshold} & $10^{-6}$ & ADAPT stops if the change in the energy is less than the threshold over ``slow\_stopper\_patience'' layers \\
\texttt{slow\_stopper\_patience} & $5$ & Number of layers to wait before stopping, if the change in energy across these layers is less than ``slow\_stopper\_threshold'' \\
\texttt{gradient\_threshold} & $10^{-6}$ & Minimum gradient norm for an operator to be considered as a candidate in the mixer layer \\
\texttt{score\_stopper\_threshold} & $10^{-6}$ & Stops the algorithm if all of the operators in the pool have a gradient (score) less than this threshold \\
\texttt{parameter\_stopper\_max} & $200$ & Maximum number of variational parameters allowed before the algorithm stops \\
\bottomrule
\end{tabular}
\caption[Hyperparameters used to evaluate the performance of ADAPT-QAOA, TETRIS-QAOA and MosaicADAPT-QAOA]{Hyperparameters used to evaluate the performance of ADAPT-QAOA, TETRIS-QAOA and MosaicADAPT-QAOA on the benchmark set of Max-E3-SAT instances.}
\label{tab:hyperparams_results}
\end{table*}

\newpage
\bibliographystyle{IEEEtran}
\bibliography{refs,ilya-biblio}
\end{document}